\documentclass[%
 reprint,
 superscriptaddress,
nofootinbib,
 amsmath,
 amssymb,
 prx,
]{revtex4-2}

\usepackage{graphicx}
\usepackage{dcolumn}
\usepackage{bm}
\usepackage[colorlinks=true, linkcolor=blue, citecolor=blue, urlcolor=blue]{hyperref}
\usepackage{siunitx}
\usepackage{mathrsfs}
\usepackage{braket}
\usepackage[dvipsnames]{xcolor}
\usepackage{acronym}
\usepackage{mathtools}
\usepackage{amsmath}
\usepackage{placeins} 

\begin{document}

\newacro{spdc}[SPDC]{spontaneous parametric down-conversion}
\newacro{bbo}[BBO]{Barium Borate}
\newacro{bibo}[BiBO]{Bismuth Borate}
\newacro{jsa}[JSA]{joint spectral amplitude}
\newacro{apd}[APD]{avalanche photodiode}
\newacro{hom}[HOM]{Hong, Ou, and Mandel}
\newacro{qd}[QD]{quantum dot}

\preprint{APS/123-QED}

\title{Symmetry allows for distinguishability\texorpdfstring{\\}{} in totally destructive many-particle interference}

\author{Julian Münzberg}
\email{julian.muenzberg@uibk.ac.at}
\affiliation{Institut für Experimentalphysik, Universität Innsbruck, Technikerstr.~25, 6020 Innsbruck, Austria}

\author{Christoph Dittel}
\affiliation{Physikalisches Institut, Albert-Ludwigs-Universität Freiburg, Hermann-Herder-Str.~3, 79104 Freiburg, Germany}
\affiliation{EUCOR Centre for Quantum Science and Quantum Computing, Albert-Ludwigs-Universität Freiburg, Hermann-Herder-Str.~3, 79104 Freiburg, Germany}

\author{Maxime Lebugle}
\affiliation{Eulitha, Studacherstrasse 7b, 5416 Kirchdorf, Switzerland}

\author{Andreas Buchleitner}
\affiliation{Physikalisches Institut, Albert-Ludwigs-Universität Freiburg, Hermann-Herder-Str.~3, 79104 Freiburg, Germany}
\affiliation{EUCOR Centre for Quantum Science and Quantum Computing, Albert-Ludwigs-Universität Freiburg, Hermann-Herder-Str.~3, 79104 Freiburg, Germany}

\author{Alexander Szameit}
\affiliation{Institut für Physik, University of Rostock, Albert-Einstein-Str.~23, 18059 Rostock, Germany}

\author{Gregor Weihs}
\author{Robert Keil}
\affiliation{Institut für Experimentalphysik, Universität Innsbruck, Technikerstr.~25, 6020 Innsbruck, Austria}

\begin{abstract}

We investigate, in a four photon interference experiment in a laser-written waveguide structure, how symmetries control the suppression of many-body output events of a $J_x$ unitary. 
We show that totally destructive interference does not require mutual indistinguishability between all, but only between symmetrically paired particles, in agreement with recent theoretical predictions. 
The outcome of the experiment is well described by a quantitative simulation which 
accounts for higher order emission of the photon source, imbalances in the scattering network, partial distinguishability, and photon loss.
\end{abstract}

\maketitle


\section{\label{sec:introduction}Introduction}

Many-particle interference lies at the very heart of many quantum information and computations schemes with photons \cite{Knill2001, Bouwmeester1997, Hensen2015}, since coherent superpositions of many-particle states can accommodate a level of complexity which is out of reach for deterministic classical simulations 
\cite{Broome2013,Spring2013,Tillmann2013,Crespi2013,Wang2017,Paesani2019,Wang2019,Zhong2020,Tichy2010,Arute2019,Giordani2018,Walschaers2016,Pednault2019}. A necessary condition 
for quantum interference on the many-particle level is the indistinguishability of distinct many-particle transition amplitudes, thus lifting wave particle duality \cite{Englert1996} from the single particle to the 
many particle level \cite{dittel2019wave}. 

Importantly, such indistinguishability of amplitudes relies on {\em two} distinct aspects of the specific physical setting -- the superimposed amplitudes' topologies, and the indistinguishabilities of the involved particles. The topology, determined by the external potential seen by the particles, determines the actual number of particles which potentially populate the contributing many-particle transition amplitudes, and thereby defines the subsets of particles which ought to be mutually indistinguishable for many-particle interference to impact on the experimentally accessible counting statistics. 

In the paradigmatic setting of two-particle interference due to \ac{hom} \cite{Hong1987}, where two photons, with controllable degree of \mbox{(in-)}distinguishability, each enter one of the input modes of a balanced beamsplitter, the topological aspect remains trivial, since only two two-particle transition amplitudes (both photons reflected, or both photons transmitted at one single, symmetric potential barrier) are coupled, and the experimental phenomenology thus entirely hinges on the indistinguishability of the incoming photons. Consequently, the famous strict suppression of the coincident output event, with one particle detected in each output port, reliably certifies that indistinguishability.

A generalisation of the \ac{hom} setting to larger numbers of modes and particles is possible, giving rise, under suitable symmetry requirements for the (single-particle) unitary which defines the non-interacting many-particle scattering process, to so-called suppression laws \cite{Lim_2005,Tichy2010,Tichy2012,Perez-Leija2013,Crespi2015,Dittel2017a} 
which identify those transition events which are strictly suppressed by destructive many-particle interference. These have been validated 
experimentally \cite{Spagnolo2013,Carolan2015,Crespi2016,Su2017,Weimann2016a,Viggianiello2018,Leedumrongwatthanakun2020,Ehrhardt2020}, 
and found a theoretical formulation \cite{Dittel2018, Dittel2018a} which -- by algebraic considerations, hence applicable for arbitrary system sizes -- anchors them to the 
symmetry properties of the many-particle input state and of the scattering unitary under permutations. Since symmetry considerations reduce the complexity of the 
general many-particle interference problem, the experimental validation of said suppression rules was suggested as a viable certification protocol for bona-fide many-particle interference 
phenomena, as it directly assesses the granular features of many-particle quantum interference, in contrast, e.g., to mean-field samplers \cite{Tichy2014,Crespi2016, Viggianiello2018,Viggianiello2018a,Walschaers2020}. 

However, because of the, in general, non-trivial imprint of the transition amplitudes' topology it is important to realise that many-body interference effects do not necessarily require mutual indistinguishability of all involved particles \cite{Jones2020}. Specifically for many-particle suppression rules, this implies that many-particle transmission events can be perfectly suppressed even for pairwise perfectly distinguishable subsets of particles, if only the topology of the superimposed transition amplitudes remains invariant under their exchange, which is ultimately controlled by the symmetry properties of the injected many-particle state
\cite{Dittel2019}.    

Here we experimentally unfold this refinement of the physics underlying many-particle suppression laws, by implementing the 
evolution of four photons under the so-called $J_x$ unitary transformation, for different mutual distinguishabilities between the particles. So far, this scattering scenario has only been investigated experimentally with photon pairs, 
making it impossible to distinguish between the effect of symmetry and full indistinguishability \cite{Perez-Leija2013,Weimann2016a}. However, by using four photons we here show that the suppression effect persists as long as the initial state exhibits the required symmetry, even if two pairs of indistinguishable particles are made fully distinguishable. Our findings are in agreement with recent theoretical insights and open new perspectives in the characterization and validation of many-body indistinguishability.

This work is structured as follows: In Sec.~\ref{sec:suppression_law} we recap the suppression law for the $J_x$ unitary and introduce the experimentally investigated input states differing by the particles' indistinguishability. Depending on these states, the scenario is expected to show either suppression or no suppression. Section \ref{sec:setup} describes the experimental setup and outlines the measurement procedure. We present the experimental results in Sec.~\ref{sec:results} and conclude in Sec.~\ref{sec:conclusion}. 
Details on further aspects of the measurement procedure and the theoretical model describing our experiment are deferred to the Appendix. 

\section{\label{sec:suppression_law} Suppression law for the \texorpdfstring{$J_x$}{Jx} unitary}

Let us consider a set of $N$ bosons initially 
prepared in a multi-mode Fock state with respect to the spatial input modes of an optical scattering device. The latter implements a unitary transformation which redistributes the particles over its $n$ modes, with the many-particle output state analysed by projective measurements of the individual modes' occupation numbers. Given that the unitary is invariant under mode permutations, and that the input state satisfies a related
permutation-symmetry, the general suppression laws formulated in \cite{Dittel2018, Dittel2018a} specify which input-output combinations
are suppressed by totally destructive many-particle interference.

The so-called $J_x$ unitary satisfies such permutation symmetry, and, according to the general suppression laws, exhibits a large number of suppressed output events for input states which  
are mirror symmetric with respect to the central mode. 
While a detailed derivation of the $J_x$ suppression law can be found in \cite{Dittel2018a, Dittel2019}, in the following we summarize the main ingredients relevant for our work. 

The single-particle $J_x$ unitary $U^\mathrm{J}(t)=e^{\mathrm{i} J_x t/\hbar}$ 
is generated by the angular momentum operator $J_x$ in $x$ direction with matrix elements \cite{Perez-Leija2013, Perez-Leija2013a, Weimann2016a} 
\begin{equation}
    \left[J_x\right]_{k,j}=\frac{\hbar}{2}\left(\sqrt{k(n-k)}\delta_{j,k+1}+\sqrt{j(n-j)}\delta_{j,k-1}\right),
    \label{eq:jx_matrix_elements}
\end{equation}
where $\delta_{j,k}$ is the Kronecker delta.
For an evolution time $t=\pi/2$, the resulting unitary $U^\mathrm{J}(\pi/2)\equiv U^\mathrm{J}$ appears, up to a phase factor, mirror symmetric with respect to the central mode \cite{Dittel2018a, Dittel2019},
\begin{equation}
    U^\mathrm{J}_{k, \pi^\mathrm{J}(j)}=U^\mathrm{J}_{k,j}\exp\left(\mathrm{i}\pi\left[k-j+\frac{n-1}{2}\right]\right),
    \label{eq:phase_relation}
\end{equation}
with
\begin{equation}
    \pi^\mathrm{J}(j)=n+1-j
    \label{eq:mode_permutation}
\end{equation}
the mirror-symmetric permutation of modes $j\in\{1,\dots,n\}$, and $\mathscr{P}^\mathrm{J}$ the corresponding single-particle permutation operator.\footnote{Note that in Eq.~\eqref{eq:phase_relation}, we attribute the columns and rows of $U^{\mathrm{J}}$ to the input and output modes, respectively. In the literature this is sometimes assumed the other way around.}
Given that the unitary evolution and the measurement of the particles' output mode occupation only act upon their {\em external} degrees of freedom -- i.e., their mode indices -- 
we can trace out all remaining -- \emph{internal} -- degrees of freedom which, potentially, render them partially distinguishable. Note that in our photonic setting, the internal degrees of freedom include, e.g., the photons' polarization, spectral properties, and arrival times. 
This procedure results in the reduced external $N$-particle density operator $\rho_\mathrm{E}$ (see \cite{Dittel2019, dittel2019wave} and Appendix~\ref{sec:appendix:model} for details).
For all mirror-symmetric input states with uncorrelated internal degrees of freedom (see Appendices \ref{sec:appendix:model_spdc} and \ref{sec:appendix:pairs_part_dist_spdc}) and an external state $\rho_\mathrm{E}$ satisfying
\begin{equation} 
[(\mathscr{P}^\mathrm{J})^{\otimes N},\rho_\mathrm{E}]=0,
\label{eq:InputSymmetry}
\end{equation}
we retrieve the following suppression law:
\textit{All output states with an odd number of particles in even output modes are suppressed} \cite{Dittel2018a}. 

Note that Eq.~\eqref{eq:InputSymmetry} does not necessarily require the mutual indistinguishability of all constituent particles. In the case of internal product states only particles occupying 
modes which belong to the same cycle $c$ of $\pi^\mathrm{J}$ -- and thus define those subsets of particles which mutually interfere to induce the suppression of the output events predicted  
by \eqref{eq:InputSymmetry} -- must share the same internal state for \eqref{eq:InputSymmetry} to hold.\footnote{In other terms, the cycle structure underlying a given input state's permutation symmetry defines the orders -- two, three, $\ldots\, N$ particles -- of multi-particle interference contributions to the predicted event suppression.}

In our experiment, we demonstrate the relation between the input state's symmetry [Eq.~\eqref{eq:InputSymmetry}] and the suppression of many-body output events for the interference of $N=4$ photons on a $J_x$ unitary with $n=7$ modes. In cycle notation, the mode permutation \eqref{eq:mode_permutation} then reads $\pi^\mathrm{J}=(1\,7)(2\,6)(3\,5)(4)$, with cycles $c_1=(1\,7)$, $c_2=(2\,6)$, $c_3=(3\,5)$, and $c_4=(4)$.
\begin{figure}
\includegraphics[width=\linewidth]{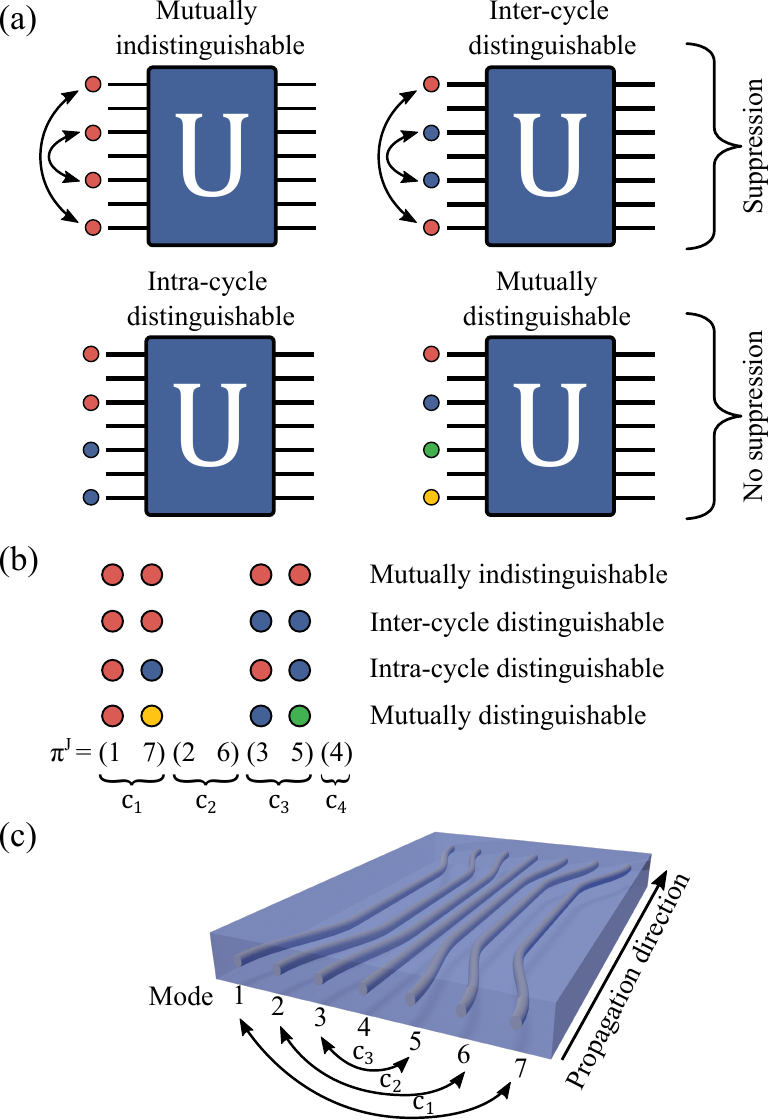}
\caption{\label{fig:permutation}
Experimentally investigated input states of the $J_x$ unitary. (a) Photons are input to the odd modes of the unitary. Four different symmetry scenarios with respect to the mirror axis (central mode) are illustrated. Suppression is expected only for the upper two scenarios. The particles' internal states are illustrated by their coloring.
In (b) the occupation of the permutation cycles $c_j$ is shown. The photonic waveguide structure implementing the $J_x$ unitary is illustrated in (c).
}
\end{figure}
We investigate four different cases of the particles' indistinguishability structure illustrated in Fig.~\ref{fig:permutation}(a). 
These scenarios correspond to the particles 
on input being {\em mutually indistinguishable}, {\em inter-cycle distinguishable}, {\em intra-cycle distinguishable}, and {\em mutually distinguishable}.
Even though the second scenario features distinguishable particles in modes belonging to distinct cycles, it remains invariant under $\pi^\mathrm{J}$, and, thus, gives rise to the above suppression law [compare Fig.~\ref{fig:permutation}(b)]. In contrast, the latter two input conformations cannot give rise to the suppression effect here under scrutiny.
The counting statistics observed in our experimental analysis unambiguously reveals the symmetry-induced minimal requirements on the mutual indistinguishability only of sub-sets of the incoming N-particle state for interference-induced output event suppression.

\section{\label{sec:setup} Experimental setup}

\begin{figure*}
\includegraphics[width=\linewidth]{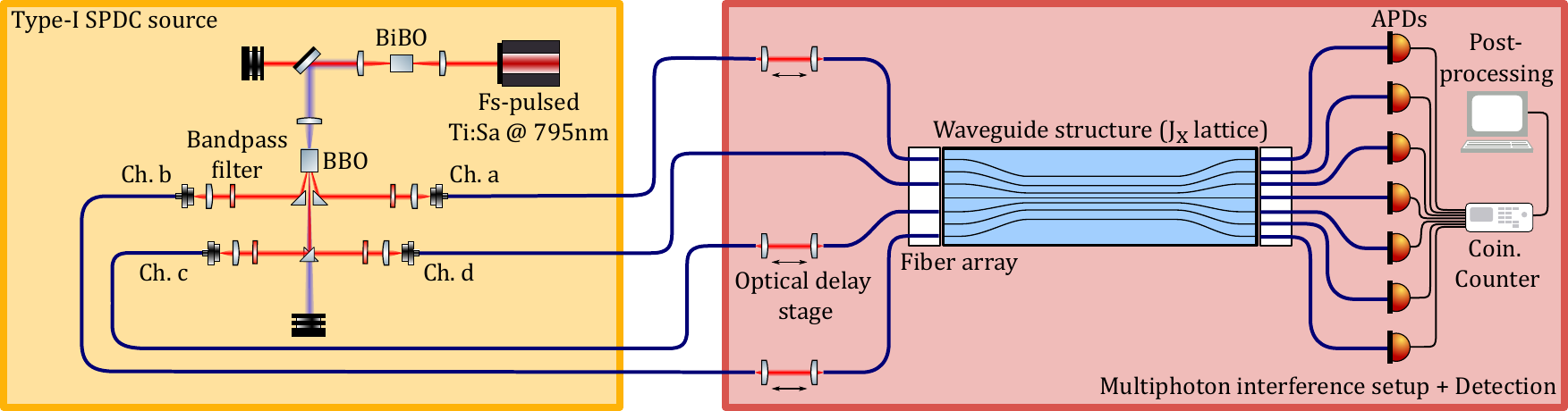}
\caption{\label{fig:setup} Schematic illustration of the four-photon \ac{spdc} source, of 
the multiphoton interference, and of the detection setup. A pulsed Ti:Sa laser was frequency doubled by second harmonic generation in a \ac{bibo} crystal in collinear configuration. The residual pump light was filtered by dichroic mirrors, while the up-converted light was focused into a \ac{bbo} crystal, where pairs of photons were generated in a type-I \ac{spdc} process. The emitted and spectrally filtered photon pairs were collected via four polarization maintaining single mode fibers labelled as channel (Ch.) a-d. Three freespace optical delay stages were used to adjust the temporal delay between photons from different channels. The light was coupled to the waveguide chip via a fiber array. The output of the waveguide structure was collected with multi-mode fibers and measured using \acp{apd} and a coincidence counter (Coin.~Counter). 
}
\end{figure*}
The many-particle interference experiment was performed on an integrated linear optics platform. 
For the implementation of the $J_x$ unitary, 
we used an array of seven evanescently coupled 
modes in a $\si{fs}$ laser-written waveguide structure in fused-silica \cite{Szameit2010, Meany2015}. 
The evanescent coupling between the waveguides was chosen according to the nearest neighbor coupling structure from Eq.~\eqref{eq:jx_matrix_elements}, with an interaction length corresponding to an evolution time $t=\pi/2$.
A schematic representation is shown in Fig.~\ref{fig:permutation}(c). See Appendix \ref{sec:appendix:unitary_reconstruction} for details on the waveguide fabrication.

The four-photon input state was generated by double-pair emission of a type-I \acf{spdc} source \cite{Ou1999}. The experimental setup is illustrated in Fig.~\ref{fig:setup}. 
A \SI{200}{fs}-pulsed Ti:Sa laser with a repetition rate of $\SI{76}{MHz}$ and a center wavelength of \SI{795}{nm} was frequency doubled in a \acf{bibo} crystal. The resulting up-converted V-polarized light with an optical power of $\approx\SI{150}{mW}$ was focused into a \acf{bbo} crystal, where photon pairs were created in a type-I non-collinear \ac{spdc} process.
The emitted H-polarized photon pairs at a center wavelength of $\SI{795}{nm}$ were collected into four polarization maintaining single-mode fibers after spectral filtering using $\approx\SI{3}{nm}$ FWHM bandpass filters, which reduces spectral correlations and improves the indistinguishability between photons from different pairs. 
The four fibers, labelled channels a-d, were oriented such that the channel pairs a and b, as well as c and d (also called mode pairs), collect photons from opposite spots of the \ac{spdc} emission cone. Hence, by momentum conservation, each mode pair collects both photons originating from the same \ac{spdc} event.
The collection of two photon pairs within their coherence time (i.e.~temporally overlapping wave packets) can arise from two different types of events:
either the photon pairs were collected by the same mode pair, e.g.~both channel a and b collect two photons, or by different mode pairs, such that each channel a-d collects one photon.

In the experiment, we connected channels a and b to input modes 1 and 7 of the $J_x$ unitary (i.e. to the modes of cycle $c_1$) and channels c and d to modes 3 and 5 (of cycle $c_3$), respectively. This connectivity was used for all four investigated scenarios. Hence, the photon collection procedure resulted in the input mode occupations
\begin{subequations}
\label{eq:input_mode_occupations}
\begin{eqnarray} 
\vec{R}_1 &= (2,0,0,0,0,0,2),\\
\vec{R}_2 &= (1,0,1,0,1,0,1),\\
\vec{R}_3 &= (0,0,2,0,2,0,0),
\end{eqnarray}
\end{subequations}
with $\vec{R}_j$ listing the number of photons in each input mode. Note that, in the ideal case of perfect indistinguishability between all particles, 
the states corresponding to all three possible input configurations satisfy the required mirror symmetry [see Eq.~\eqref{eq:mode_permutation}]. In the experiment, however, there are residual spectral correlations, rendering photons from different pairs partially distinguishable \cite{Grice1997}. 
Considering this, the described connectivity also ensures that photons from the same \ac{spdc} event always occupy modes belonging to the same cycle, such that the symmetry conditions are best satisfied in the presence of experimental imperfections.

To control the temporal delays between the photons and, thus, their mutual distinguishability, we used free-space optical delay stages in three of the four channels. The four scenarios shown in Fig.~\ref{fig:permutation}(a) were implemented by appropriately adjusting the photons' time delays. 
For coupling light into the quasi-transverse-magnetic mode of the waveguide structure and collecting the outcome, we used a commercially available polarization maintaining and multi-mode V-groove fiber array, respectively. The fibers of the output array were connected to seven \acfp{apd} with an average detection efficiency of 65\%. 

After having optimized the temporal overlap of photons from different channels in 
\ac{hom}-type measurements [see Appendix \ref{sec:appendix:measurement_procedure} for details], 
we recorded all fourfold coincidence events with a time tagging device. Only non-bunching output events (at most one photon ends up in an output mode) are experimentally accessible,
since the used \acp{apd} are non-number-resolving.
In parallel to fourfold coincidences, we collected single channel count rates, as well as twofold and threefold coincidences between all output modes [see Appendices \ref{sec:appendix:phase_fluctations} and \ref{sec:appendix:hom_vis}].

\section{Results}
\label{sec:results}

\begin{figure*}
\includegraphics[width=\linewidth]{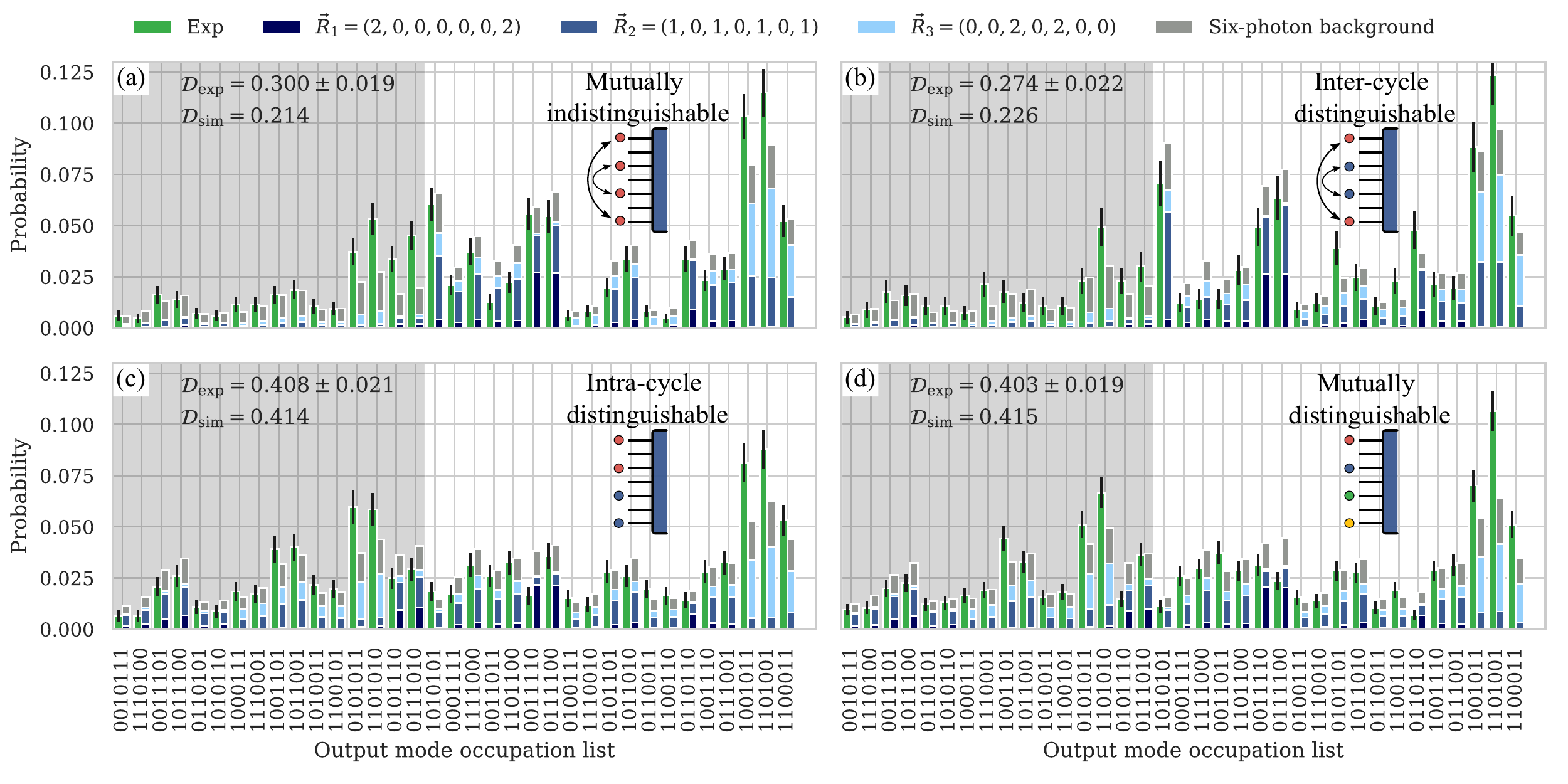}
\caption{\label{fig:results} 
Experimental and simulated output statistics of all fourfold coincidences between distinct output modes. Panel (a) and (b) show the output statistics obtained for a \emph{mutually indistinguishable} and 
for an \emph{inter-cycle distinguishable} input state, respectively, for which all output events in the grey shaded area are ideally suppressed. The output statistics obtained for an \emph{intra-cycle distinguishable} and for a \emph{mutually distinguishable} input state is shown in (c) and (d), respectively. Here no suppression effect is expected, even in the ideal case. Green bars correspond to experimental (Exp) data, 
and the error bars indicate one standard deviation of the Poissonian counting statistics. The bars to the right of the experimental data correspond to a simulation of the experiment, with the bars' dark blue, blue, light blue, and grey part indicating the contribution stemming from the input mode occupation $\vec{R}_1$, $\vec{R}_2$, $\vec{R}_3$ [see Eq.~\eqref{eq:input_mode_occupations}], and from the six-photon background of the photon source, respectively. 
}
\end{figure*}

For each scenario shown in Fig.~\ref{fig:permutation}(a), we recorded fourfold coincidences between distinct output modes. Dependent on the scenario, the average event rate 
ranges between 0.0023 and \SI{0.0045}{Hz}, with a total number of collected events, $N_\mathrm{total}$, between 566 and 1154 [see Tab.~\ref{tab:results}]. The experimental results are shown in Fig.~\ref{fig:results} together with the theoretical predictions obtained from a simulation of the experiment. Experimental and simulated results are both normalized.
In Fig.~\ref{fig:results}(a) and (b), all output events in the grey shaded area 
are suppressed in the ideal case, as predicted by the suppression law.
While our data is substantially biased by higher-order multi-photon-pair emission (grey bars in Fig.~\ref{fig:results}; $\approx 34\%$ of all registered four-photon events can be attributed to the six-photon background), we can confirm a clear reduction of all ideally suppressed events in the case of \emph{mutually indistinguishable} and \emph{inter-cycle distinguishable} particles [see Fig.~\ref{fig:results}(a)~and~(b)], to be contrasted with 
the cases of \emph{intra-cycle} and \emph{mutually distinguishable} particles [see Fig.~\ref{fig:results}(c)~and~(d)], where no suppression effect is expected nor observed. 

In order to benchmark the suppression, we use the degree of suppression violation 
$\mathcal{D}=N_{\mathrm{forbidden}}/N_{\mathrm{total}}$ \cite{Viggianiello2018} with $N_{\mathrm{forbidden}}$ the number of recorded four-photon coincidences corresponding to ideally suppressed output events, and $N_{\mathrm{total}}$ the total number of recorded events. Note that in the ideal case $\mathcal{D}=0$. However, we expect a non-vanishing value of $\mathcal{D}$ due to fabrication errors of the waveguide structure, imperfections in the photons' indistinguishability (caused by their spectro-temporal properties), and higher-order multi-pair emission. 

In our theoretical simulation of the experiment we account for the above experimental imperfections. Pertinent to photon pairs from \ac{spdc} are their spectral correlations arising from energy and momentum conservation in the generation process, such that the combined internal state of all particles does not factorize into individual internal states of the particles \cite{Grice1997}. Not only does this spectral entanglement influence the ratio between spontaneous and stimulated emission in the \ac{spdc} process and, thereby, the photon number distribution \cite{Ou1999a,DeRiedmatten2004,Ou2006}, it also strongly affects the many-particle interference for configurations with more than one photon occupying the same initial mode. The established method of calculation using the so-called partial indistinguishability matrix \cite{Shchesnovich2015} is applicable to this situation, but computationally costly, as it requires a double summation over all permutations of particle orderings. The more efficient formalism from Ref.~\cite{Tichy2015}, however, cannot be used here as it relies on factorizable internal states.
Adopting the formalism of \cite{Dittel2019}, we here derive a new framework for many-particle interference of correlated photon pairs, which avoids the costly summation over all particle orderings and replaces it by a faster summation over \textit{inequivalent} orderings only. This framework is the first, to our knowledge, which simultaneously incorporates correlated internal states, multiple occupancy of external input modes, as well as the correct weights between stimulated and spontaneous emission in the various multi-pair contributions of the \ac{spdc}, and is described in detail in Appendix \ref{sec:appendix:model}.

The experimentally measured and theoretically predicted degrees of suppression violation are summarized in Table \ref{tab:results}.
\begin{table}
\caption{\label{tab:results}
For the four investigated scenarios illustrated in Fig~\ref{fig:permutation}(a), the measured ($\mathcal{D}_\mathrm{exp}$) and simulated ($\mathcal{D}_\mathrm{th}$) degree of suppression violation is listed together with the total number $N_\mathrm{total}$ of recorded four-photon coincidences. 
}
\begin{ruledtabular}
\begin{tabular}{llll}
        & $N_\mathrm{total}$ & $\mathcal{D}_\mathrm{exp}$ & $\mathcal{D}_\mathrm{th}$\\
       \colrule
        \emph{mutually indistinguishable} & 862 & $0.300\pm0.019$ & 0.214\\
        \emph{inter-cycle distinguishable} & 566 & $0.274\pm0.022$ & 0.226\\
        \emph{intra-cycle distinguishable} & 921 & $0.408\pm0.021$ & 0.414\\
        \emph{mutually distinguishable} & 1154 & $0.403\pm0.019$ & 0.415\\
\end{tabular}
\end{ruledtabular}
\end{table}
As quantified by the degree of suppression violation for both \emph{mutually indistinguishable} and \emph{inter-cycle distinguishable} particles there is a significantly smaller fraction of events in the grey area of Fig.~\ref{fig:results} compared to the cases of \emph{mutually distinguishable} and \emph{intra-cycle distinguishable} particles which violate the mirror symmetry~\eqref{eq:InputSymmetry}. This confirms the presence of destructive many-particle interference in the two symmetric scenarios alone.

Comparison of theoretical and experimental results [see Fig.~\ref{fig:results}] yields reasonable agreement. Yet, for the cases \textit{mutually indistinguishable} and \textit{inter-cycle distinguishable} [see Fig.~\ref{fig:results}(a) and (b)], the simulation predicts a significantly lower degree of violation than measured. We attribute these deviations mainly to the following systematic errors and limitations of the model: 
1) The experimental unitary reconstruction is subject to
imprecisions, and the theoretically assumed phase relations may not be perfectly realised 
[see Appendix~\ref{sec:appendix:unitary_reconstruction}].
2) In addition to the substantial contributions from six generated photons on the statistics of four-photon outputs, there is also an eight-photon background (i.e.~quadruple pair productions) that we estimate to comprise about 5\% of all detected events. This background is not accounted for, due to the computational overhead.
However, if we assume a uniform distribution of this eight-photon background among all fourfold outputs (suppressed and unsuppressed), we estimate that $\mathcal{D}_\mathrm{th}$ would increase from initially 0.214 to 0.226 for the \emph{mutually indistinguishable} scenario. 
3) Inaccuracies in the estimated generation probabilities and losses before the unitary would lead to a different weighting of the contributions of the input states. That in turn 
may shift the probability distribution of the output events.
4) Finally, the indistinguishabilties of photons from the same pair, as well as of photons from different pairs, are estimated via 
\ac{hom} measurements [see Appendix~\ref{sec:appendix:source_setup_parameters}]. We noticed a drift in the indistinguishability during the measurement, which we correct for in the simulation, but this correction has its own precision limits [see Appendix~\ref{sec:appendix:hom_vis}]. Errors in the estimated indistinguishabilities lead to errors in the probability distribution of output events.

The optimal degree of suppression achievable in the experiment is mostly
limited by the unwanted multi-photon background of the photon source. From a simulation excluding the six-photon background, we retrieve a theoretically predicted degree of suppression violation of 0.102 (compared to 0.214 with background) for the \textit{mutually indistinguishable} case. It is possible to reduce this background by decreasing the \ac{spdc} pump power, which, however, also decreases the source brightness, leading to a longer integration time 
to achieve the same experimental quality of the output statistics. Since
our integration time per configuration was 3-4 days, this is only possible if one improves the transmission efficiency at the same time, e.g.~via enhanced coupling from the fiber to the chip.
Alternatively, one can replace the non-deterministic \ac{spdc} source by a quasi-deterministic source e.g.~by actively demultiplexing photons from a semiconductor \ac{qd} \cite{Mendoza2016, Lenzini2017, Wang2017, Hummel2019a}. Resonantly excited semiconductor \acp{qd} feature almost zero multi-photon emission \cite{Schweickert2018, Hanschke2018, Scholl2019} and higher source brightness. Hence, they can be used to further optimise the suppression and,
at the same time, increase the precision of the experiment.

\section{\label{sec:conclusion} Conclusion}

In this work, we experimentally demonstrated that totally destructive many-particle interference (aka event suppression on output) does not necessarily require 
mutual indistinguishability of all involved particles. 
Instead, merely the input state's symmetry matters, such that only particles occupying modes belonging to the same cycle of the underlying permutation-symmetry must be indistinguishable.
As a figure of merit, we used the degree of suppression violation $\mathcal{D}_{\mathrm{exp}}$, which is significantly smaller for input states satisfying the symmetry required for the suppression as compared to input states violating this symmetry (and, thereby, destroying the necessary indistinguishability of those particles populating interfering many-particle amplitudes which, by the input state's cycle structure, contribute to the predicted suppression event).
In order to simulate our experiment we developed a general theoretical model designed to describe many-particle interference of photons with arbitrarily correlated spectral properties. Our model allows for the multifold occupation of input modes, incorporates quantitatively correct weights between the various emission orders in realistic \ac{spdc} photon sources with imperfections, and predicts the experimentally observed four-photon statistics with reasonable accuracy. 

Our results clearly demonstrate that the observation of many-body suppression does not necessarily suffice to conclude that all involved constituents are mutually 
indistinguishable. 
This refines the suggestion that fully destructive many-particle interference across highly symmetric scattering devices can unambiguously certify a boson sampling device: In general, the output event suppression by multi-particle interference only certifies the symmetry of the input state with respect to a permutation of modes, and the indistinguishability of those particles pertaining to subgroups identified by the diverse cycles of that very permutation. Only when this permutation consists of a single cycle of length $N$, as in the case of the $N\times N$ Fourier unitary \cite{Tichy2012,Dittel2019}, 
does suppression suffice to certify the particles' mutual indistinguishability. 

\begin{acknowledgments}
We acknowledge useful discussion with Stefan Frick. 
G.W., R.K., and J.M.~acknowledge support from the Austrian Science Fund (FWF Projects No.~I2562, P30459, and F7114).
A.S.~thanks the Deutsche Forschungsgemeinschaft (Grants SZ 276/9-2, SZ 276/12-1, SZ 276/20-1, SZ 276/21-1, BL 574/13-1) and the Alfried Krupp von Bohlen and Halbach Foundation for financial support. M.L.~acknowledges funding from the Marie Curie Actions within the Seventh Framework Programme for Research of the European Commission, under the Initial Training Network PICQUE, Grant No.~608062.
C.D.~acknowledges the Georg H.~Endress foundation for financial support.
\end{acknowledgments}

\appendix

\section{\label{sec:appendix:model} Many-particle interference from SPDC photons}

In this section, we present our theoretical model to describe many-particle interference experiments with photon pairs collected from a probabilistic multi-pair \ac{spdc} source. 
Note that our model is not limited to the present setup, but can be used in a wide variety of experimental settings to calculate the input-output probabilities of pairwise correlated and partially distinguishable multi-photon states under any unitary transformation $U$.
Similar considerations have been made for two photons fed into a Sylvester interferometer \cite{Viggianiello2018}, where the authors consider fabrication errors in the unitary, multiphoton emission, and partial distinguishability of the photons.
Here, our description generalizes this approach to multiple partially distinguishable photon pairs.

The model is divided into the following steps: We construct the $N$-photon input state (with, in general, $N>4$) consisting of $N/2$ photon pairs obtained from \ac{spdc} and calculate the generation probabilities of these states in Sec.~\ref{sec:appendix:model_spdc}. The particles' internal states are calculated from the product of reconstructed \acp{jsa} of the emitted photon pairs. Loss before the unitary is modeled by extending the optical modes of the unitary with additional ancillary modes and unbalanced beamsplitters, with lost photons being coupled into these ancillary modes [see Sec.  \ref{sec:appendix:loss_model}]. 
Afterwards, for each so-constructed and loss-weighted input state, input-output probabilities are calculated according to our formalism presented in Sec.~\ref{sec:appendix:pairs_part_dist_spdc}. This requires the matrix representation of the many-particle density operator, 
which accounts for partial distinguishability of the photons as calculated from the overlaps of the permuted internal states from Sec.~\ref{sec:appendix:model_spdc}.

In general, the relative phases between all input states must be taken into account in the calculation of the output probabilities, since these phases result in an interference of the probability amplitudes corresponding to the different input states \cite{Tichy2011,Carolan2014}. However, observing the time traces of recorded two-fold coincidence counts shows that these relative phases fluctuate [see Appendix~\ref{sec:appendix:phase_fluctations}] much faster than the integration time of the experiments. Therefore, the experimental data results from an incoherent mixture of these input states, such that we can consider an incoherent superposition of the input-output probabilities of all contributing input states weighted by their generation probabilities and losses.
The resulting probabilities are then renormalized, accounting for the relative output transmissivities and detector efficiencies, which were obtained in the unitary reconstruction process [see Appendix \ref{sec:appendix:unitary_reconstruction}].

\subsection{\label{sec:appendix:model_spdc} Model of the SPDC source}

We consider the four-mode \ac{spdc} source [cf. Fig.~\ref{fig:setup}] as two independent, pulsed pair sources, which produce two quantum states $\ket{\Psi^{(\mathrm{ab})}}$ and $\ket{\Psi^{(\mathrm{cd})}}$ in the two mode pairs of interest. The overall state reads \cite{Ou1999a}
\begin{equation}
    \ket{\Psi}=\ket{\Psi^{(\mathrm{ab})}}\otimes\ket{\Psi^{(\mathrm{cd})}}.
\end{equation}
The state of each parametric process can be written as a sum over the emission of $P$ photon pairs, with the state generated by the first source being \cite{Ou2006}
\begin{equation}
\ket{\Psi^{(\mathrm{ab})}} = c^{(\mathrm{ab})}\sum_{P=0}^{\infty}\frac{\left(p^{(\mathrm{ab})}\right)^{P/2}}{P!}\sqrt{\mathcal{N}^{(\mathrm{ab})}_P}\ket{\Psi^{(\mathrm{ab})}_P}.
\label{eq:multiphoton_state}
\end{equation}
Here, $c^{(\mathrm{ab})}\lesssim 1$ is a normalization constant and $\left(c^{(\mathrm{ab})}\right)^2p^{(\mathrm{ab})}\approx p^{(\mathrm{ab})}$ is the pair generation probability per pump pulse.
The state of $P$ photon pairs produced in a pump pulse is given in second quantization by an integral over the particles' internal states, i.e.~their frequencies \cite{Ou2006},
\begin{eqnarray}
& \ket{\Psi^{(\mathrm{ab})}_P} = \frac{1}{\sqrt{\mathcal{N}^{(\mathrm{ab})}_P}}\int \mathrm{d}\vec{\omega}\,\Phi^{(\mathrm{ab})}(\omega_1, \omega_1')\cdots\Phi^{(\mathrm{ab})}(\omega_P, \omega_{P}')\times\nonumber\\
    & \hat{a}_\mathrm{a}^{\dagger}(\omega_1)\hat{a}_\mathrm{b}^{\dagger}(\omega_1')\cdots \hat{a}_\mathrm{a}^{\dagger}(\omega_P)\hat{a}_\mathrm{b}^{\dagger}(\omega_P')\ket{0},
\label{eq:state-P-pairs}
\end{eqnarray}
with $\Phi^{(\mathrm{ab})}(\omega, \omega')$ the \ac{jsa} of a single photon pair, and $\mathrm{d}\vec{\mathcal{\omega}}=\mathrm{d}\omega_1\mathrm{d}\omega'_1\ldots\mathrm{d}\omega_P\mathrm{d}\omega'_P$, and $\hat{a}^\dagger_\mathrm{a}(\omega)$ the creation operator of a photon with frequency $\omega$ in mode $\mathrm{a}$. The normalization coefficient $\mathcal{N}^{(\mathrm{ab})}_P$ is dictated by the commutation rules of the ladder operators and can be calculated via a summation over all intra-mode permutations:
\begin{eqnarray}
    &\mathcal{N}_P^{(\mathrm{ab})} = \int \mathrm{d}\vec{\omega}\,\left(\Phi^{(\mathrm{ab})}(\omega_1, \omega_1')\cdots\Phi^{(\mathrm{ab})}(\omega_P, \omega_{P}')\right)^*\times\nonumber\\
    &  \sum\limits_{\pi,\sigma\in S_{P}}\Phi^{(\mathrm{ab})}(\omega_{\pi(1)}, \omega_{\sigma(1)}')\cdots\Phi^{(\mathrm{ab})}(\omega_{\pi(P)}, \omega_{\sigma(P)}').\quad
\end{eqnarray}
Here, $S_P$ is the symmetric group of $P$ elements (corresponding to the $P$ particles in each of the two modes).
Equation \eqref{eq:multiphoton_state} can be interpreted as an interpolation between a two-mode squeezed vacuum state ($\mathcal{N}^{(\mathrm{ab})}_P=P!^2$) and purely accidental multipair generation ($\mathcal{N}^{(\mathrm{ab})}_P=P!$). The former arises from an uncorrelated \ac{jsa} ($\Phi(\omega,\omega')=\phi(\omega)\phi(\omega')$) and a maximal contribution of stimulated emission, and leads to perfect, heralded \ac{hom} visibility between photons from different pairs \cite{Ou1999a}, while the latter corresponds to a maximally correlated \ac{jsa}  ($\Phi(\omega,\omega')=\delta(\omega+\omega'-\omega_\mathrm{p})\Phi(\omega,\omega')=\Phi(\omega, \omega-\omega_\mathrm{p})$ with the pump frequency $\omega_\mathrm{p}$) and purely spontaneous emission, producing zero \ac{hom} visibility. 
Note that the difference between these two extremes takes no effect for single-pair emission, i.e. $P=1$.

The probability of the first source to generate the state $\ket{\Psi^{(\mathrm{ab})}_P}$ of $P$ pairs is derived from Eq.~\eqref{eq:multiphoton_state}:
\begin{equation}
    p_P^{(\mathrm{ab})} = \left(c^{(\mathrm{ab})}\right)^2\,\frac{\left(p^{(\mathrm{ab})}\right)^P}{{P!}^2}\mathcal{N}^{(\mathrm{ab})}_P.
\end{equation}
The joint probability to generate $P$ pairs in channels a and b, and $Q$ pairs in channels c and d is then
\begin{equation}
        p_{P,Q}=\left(c^{(\mathrm{ab})}c^{(\mathrm{cd})}\right)^2 \left(p^{(\mathrm{ab})}\right)^P \left(p^{(\mathrm{cd})}\right)^Q \frac{\mathcal{N}_P^{(\mathrm{ab})}\mathcal{N}_Q^{(\mathrm{cd})}}{{P!}^2\,{Q!}^2}.
    \label{eq:gen_prob_spdc}
\end{equation}

From a preliminary characterization of the source, we obtain $p^{(\mathrm{ab})}=0.026$ and $p^{(\mathrm{cd})}=0.033$ [see Appendix \ref{sec:appendix:source_setup_parameters}]. 
In order to calculate $\mathcal{N}^{(\mathrm{ab}),(\mathrm{cd})}_P$, we construct \acp{jsa} $\Phi^{(\mathrm{ab}),(\mathrm{cd})}(\omega, \omega')$ that reproduce our experimentally measured \ac{hom} and heralded \ac{hom} visibilities 
[see Appendix \ref{sec:appendix:hom_vis}]. From this we obtain numeric values of $\mathcal{N}^{(\mathrm{ab})}_2\approx\mathcal{N}^{(\mathrm{cd})}_2\approx 3.18$ and $\mathcal{N}^{(\mathrm{ab})}_3\approx\mathcal{N}^{(\mathrm{cd})}_3\approx 21.65$. 
Note that in order to account for the distinguishability between the particles induced by different time-of-arrivals in the experimentally investigated cases of \emph{inter-cycle distinguishable} and \emph{intra-cycle distinguishable} particles, we include an additional  phase $e^{\mathrm{i}\omega\tau}$ with a sufficiently large temporal delay $\tau$ between the \ac{spdc} source channels.
With Eq.~\eqref{eq:gen_prob_spdc}, we calculate the generation probability of states up to six photons, which is summarized in Table \ref{tab:gen_prob}. We find a 32 times enhanced generation probability of four-photon states compared to six-photon states. Despite this seemingly low rate of created six-photon states, they nevertheless contribute significantly to the measured counting statistics due a combinatorial advantage over four-photon states in the lossy setup: To yield a four-photon coincidence from an initial six-photon state, up to two photons may be lost before detection. There are $\binom{6}{2}=15$ possible combinations of lost photons, which significantly increases the relative weight of six-photon states in the lossy setup. From the simulation, we predict that 33.6\% to 35.2\% (depending on the measured scenario) of all registered non-bunching output events originate from an initial six-photon creation.

\begin{table}
\caption{\label{tab:gen_prob}
All states consisting of up to six photons, that possibly lead to a fourfold coincidence after evolving under the unitary transformation $U_\mathrm{exp}$. $\vec{R}_{(\mathrm{a,b,c,d})}$ is the mode occupation list of the four \ac{spdc} channels, $N$ is the number of photons in that state, $\vec{R}_{J_x}$ is the mode occupation list for the $n=7$ input modes of the $J_x$ unitary (excluding the additional unoccupied ancillary input modes), $p_\mathrm{gen}$ is the generation probability calculated via Eq.~\eqref{eq:gen_prob_spdc}, and $p_\mathrm{gen,\,norm}$ is the generation probability normalized to all listed states. Note that despite the very low generation probabilities of the six-photon states, they contribute significantly to the final output probability due to a combinatorial advantage in the lossy setup.
}
\begin{ruledtabular}
\begin{tabular}{ccccc}
       $\vec{R}_{(\mathrm{a,b,c,d})}$ & $N$ & $\vec{R}_{J_x}$ & $p_\mathrm{gen}$ & $p_\mathrm{gen,\,norm}$\\
       \colrule
         (2, 2, 0, 0) & 4 & $\vec{R}_1$ & 0.000493 & 0.230\\
         (1, 1, 1, 1) & 4 & $\vec{R}_2$ & 0.000788 & 0.368\\
         (0, 0, 2, 2) & 4 & $\vec{R}_3$ & 0.000797 & 0.372\\
         (3, 3, 0, 0) & 6 & (3, 0, 0, 0, 0, 0, 3) & 0.000010 & 0.004\\
         (2, 2, 1, 1) & 6 & (2, 0, 1, 0, 1, 0, 2) & 0.000016 & 0.007\\
         (1, 1, 2, 2) & 6 & (1, 0, 2, 0, 2, 0, 1) & 0.000020 & 0.010\\
         (0, 0, 3, 3) & 6 & (0, 0, 3, 0, 3, 0, 0) & 0.000020 & 0.009\\
\end{tabular}
\end{ruledtabular}
\end{table}

\subsection{\label{sec:appendix:loss_model} Photon loss}

Loss processes before the fabricated $J_x$ unitary are summarized in the channel transmission factors $\eta_j$ (see Appendix \ref{sec:appendix:source_setup_parameters} for their values and the method of characterization). Their combined impact on the dynamics can be modelled by extending $U_{\mathrm{exp}}$ with seven uncoupled ancillary modes and prepending to it a unitary of seven unbalanced beamsplitters (one for each mode) feeding the lost photons into these ancillary modes with probability $1-\eta_j$. This extended unitary $\mathcal{U}$ is then used in the subsequent calculations.

\subsection{\label{sec:appendix:pairs_part_dist_spdc} Partial distinguishability of correlated photon pairs}
Let us consider a state of $N$ partially distinguishable photons, which occupy external states  $\ket{\vec{E}}=\ket{E_1,\dots, E_N}$ of an $n$-mode scattering network with unitary $\mathcal{U}$. The $N$-particle state is determined by its mode occupation list $\vec{R}=(R_1,\dots,R_n)$ or equivalently by its mode assignment list $\vec{E}=(E_1,\dots,E_N)$, with $E_\alpha$ the mode occupied by the $\alpha$th particle \cite{Dittel2018}. The particles' internal state is modelled by frequencies $\ket{\vec{\omega}}=\ket{\omega_1,\dots,\omega_N}$, yielding the combined state 
\begin{equation}
    \ket{\vec{E}}\otimes\ket{\vec{\omega}}=\ket{E_1,\dots,E_N}\otimes\ket{\omega_1,\dots,\omega_N}.
\end{equation}
The combined state of a correlated photon pair from \ac{spdc} in the mode pair a, b is described by an integral over the particles' internal states, which are correlated according to the \ac{jsa} $\Phi(\omega_1,\omega_2)$. 
In the first quantization without symmetrization, the combined state is given by
\begin{equation}
    \int\mathrm{d}\omega_1\mathrm{d}\omega_2\,\Phi(\omega_1,\omega_2)\ket{\mathrm{a,b}}\ket{\omega_1,\omega_2}.\nonumber
\end{equation}
To correctly include the exchange symmetry of bosons, the two-photon state needs to be symmetric under exchange of particles. After the required symmetrization, the state of one created photon pair reads 
\begin{equation}
\ket{\Psi_1^{(\mathrm{ab})}} = \int\mathrm{d}\omega_1\mathrm{d}\omega_2\,\Phi(\omega_1,\omega_2)\ket{\vec{R}\,:\,\vec{\omega}}
\end{equation}
with $\ket{\vec{R}\,:\,\vec{\omega}}=\left(\ket{\mathrm{a,b}}\ket{\omega_1,\omega_2}+\ket{\mathrm{b,a}}\ket{\omega_2,\omega_1}\right)/\sqrt{2}$ and the normalization condition of the \ac{jsa} is $\int\mathrm{d}\omega_1\mathrm{d}\omega_2\,\left|\Phi(\omega_1,\omega_2)\right|^2=1$.

Next, we consider a state of $P$ photon pairs emitted from the first source into mode a and b, as well as $Q$ photon pairs emitted from a second source into mode c and d. The total number of photons is $N=2(P+Q)$. We can describe the state by symmetrizing the expression 
\begin{equation}
    \int\mathrm{d}\,\vec{\omega}\Phi(\vec{\omega})\ket{\vec{E}}\ket{\vec{\omega}},\nonumber
\end{equation}
with the total \ac{jsa} given by the product of \acp{jsa} corresponding to all photon pairs,
\begin{equation}
    \Phi(\vec{\omega})=\prod_{i=1}^{P} \Phi^{(\mathrm{ab})}(\omega_{2i-1}, \omega_{2i})
    \prod_{j=P+1}^{P+Q} \Phi^{(\mathrm{cd})}(\omega_{2j-1}, \omega_{2j}),
    \label{eq:internalstate_multipairs}
\end{equation}
as well as 
\begin{equation}
    \ket{\vec{E}}=\ket{\underbrace{\mathrm{a,b}\dots,\mathrm{a,b}}_{\mathrm{a,b};\, P\, \mathrm{times}},\underbrace{\mathrm{c,d}\dots,\mathrm{c,d}}_{\mathrm{c,d};\, Q\,\mathrm{times}}},\nonumber
\end{equation}
 and $\mathrm{d}\vec{\omega}=\mathrm{d}\omega_1\dots\mathrm{d}\omega_N$. Symmetrizing the above equation yields the state
\begin{equation}
    \ket{\Psi_{P,Q}}\propto\sum_{\pi\in S_N}\ket{\vec{E}_\pi}\otimes\int\mathrm{d}\vec{\omega}\,\Phi(\vec{\omega})\ket{\vec{\omega}_\pi},
    \label{eq:spdc_state_symmetric}
\end{equation}
where we use the shorthands $\ket{\vec{E}_\pi}=\ket{E_{\pi(1)},\dots,E_{\pi(N)}}$ and $\ket{\vec{\omega}_\pi}=\ket{\omega_{\pi(1)},\dots,\omega_{\pi(N)}}$.
The calculation of Eq.~\eqref{eq:spdc_state_symmetric} can be simplified by reducing the summation over all permutations $S_N$ to a summation over permutations leading to inequivalent external states $\ket{\vec{E}_\pi}$, that is, to permutations which only permute particles across different modes. To this end, one can decompose $\pi\in S_N$ as $\pi=\xi\mu$ with $\xi\in S_{\vec{R}}$ and $\mu\in \Sigma$. Here, $S_{\vec{R}}=S_{R_1}\otimes \cdots \otimes S_{R_N}$ is a Young subgroup of $S_N$, and $\Sigma$ the right transversal of $S_{\vec{R}}$ in $S_N$ \cite{Dittel2019, dittel2019wave}. This results in
\begin{equation}
    \ket{\Psi_{P,Q}}\propto\sum_{\mu\in\Sigma,\xi\in S_{\vec{R}}}\ket{\vec{E}_{\xi\mu}}\otimes\int\mathrm{d}\vec{\omega}\,\Phi(\vec{\omega})\ket{\vec{\omega}_{\xi\mu}}.
\end{equation}
We utilize $\ket{\vec{E}_{\xi\mu}}=\ket{\vec{E}_\mu}$ and write $\vec{\omega}$ as $\vec{w}_{\xi'}$ with $\xi'\in S_{\vec{R}}$, and $\vec{w}\in X_{\vec{w}}=\{\vec{w}_\xi\, |\, \xi\in S_{\vec{R}}\}$ (such that $\{\vec{\omega}\}=\cup_{\vec{w}}X_{\vec{w}}$ and $\int\mathrm{d}\vec{\omega}=\int\mathrm{d}\vec{w}\sum_{\xi'\in S_{\vec{R}}}$).
This yields 
\begin{equation}
    \ket{\Psi_{P,Q}}
    \propto
    \sum_{\mu\in\Sigma}
    \ket{\vec{E}_{\mu}}
    \otimes
    \int\mathrm{d}\vec{w}\sum_{\xi,\xi'\in S_{\vec{R}}}\,\Phi(\vec{w}_{\xi'})\ket{\vec{w}_{\xi'\xi\mu}}.
\end{equation}
Since $S_{\vec{R}}$ forms a group, we can substitute $\xi'\xi=\xi''$ and sum over $\xi''$ 
instead of $\xi'$,
\begin{equation}
    \ket{\Psi_{P,Q}}
    \propto
    \sum_{\mu\in\Sigma}
    \ket{\vec{E}_{\mu}}
    \otimes
    \int\mathrm{d}\vec{w}\sum_{\xi,\xi''\in S_{\vec{R}}}\,\Phi(\vec{w}_{\xi''(\xi)^{-1}})\ket{\vec{w}_{\xi''\mu}}.
\end{equation}
Next, we use the same trick as above and write $\vec{w}_{\xi''}$ as $\vec{\omega}$ with $\int\mathrm{d}\vec{w}\sum_{\xi''\in S_{\vec{R}}}=\int\mathrm{d}\vec{\omega}$, and instead of summing over $\xi$, we sum over $(\xi)^{-1}$ (this substitution can be done since $S_{\vec{R}}$ forms a group), which results in 
\begin{equation}
    \ket{\Psi_{P,Q}}
    \propto
    \sum_{\mu\in\Sigma}
    \ket{\vec{E}_{\mu}}
    \otimes
    \int\mathrm{d}\vec{\omega} \left(\sum_{\xi\in S_{\vec{R}}} \Phi(\vec{\omega}_{\xi}) \right)\ket{\vec{\omega}_\mu}.
    \label{eq:app:state}
\end{equation}
We identify the not yet normalized coefficients (in the parentheses)
\begin{equation}
    \bar{C}(\vec{\omega})=\sum_{\xi\in S_{\vec{R}}} \Phi(\vec{\omega}_{\xi}).
\end{equation}
Normalizing Eq.~\eqref{eq:app:state} then leads to \cite{Dittel2019, dittel2019wave}
\begin{equation}
    \ket{\Psi_{P,Q}}=\frac{1}{\sqrt{R}}\sum_{\mu\in\Sigma}\ket{\vec{E}_\mu}\otimes\ket{\Omega_\mu},
    \label{eq:app:state_norm}
\end{equation}
with $R=N!/|S_{\vec{R}}|=N!/(P!Q!)^2$ and
\begin{equation}
    \ket{\Omega_\mu}=\int\mathrm{d}\vec{\omega}\,C(\vec{\omega})\ket{\vec{\omega}_\mu}.
\end{equation}
The internal state coefficients $C(\vec{\omega})$ take into account the possible orderings within a mode and are calculated from
\begin{equation}
        C(\vec{\omega})=\frac{\sum_{\xi\in S_{\vec{R}}}\Phi(\vec{\omega}_\xi)}{\sqrt{\int\mathrm{d}\vec{\omega'}\left|\sum_{\xi\in S_{\vec{R}}}\Phi(\vec{\omega'}_\xi)\right|^2}},
\end{equation}
with the normalization factor in the denominator such that $\int\mathrm{d}\vec{\omega}|C(\vec{\omega})|^2=1$. 

The many-particle state~\eqref{eq:app:state_norm} describes the initial state of our experiment. Since the unitary transformation $\mathcal{U}$ and the measurement of the particles' output mode occupation doesn't act upon the particles' internal degrees of freedom, we can trace them out, resulting in the reduced external many-particle state 
\begin{align}
\rho_\mathrm{E}&=\mathrm{Tr}_\mathrm{I}(\ket{\Psi_{P,Q}}\bra{\Psi_{P,Q}})\nonumber\\
&= \sum_{\mu,\nu \in \Sigma} [\rho_\mathrm{E}]_{\mu,\nu} \ket{\vec{E}_\mu} \bra{\vec{E}_\nu},
\end{align}
with
\begin{equation}
[\rho_\mathrm{E}]_{\mu,\nu}=\frac{1}{R} \langle \Omega_\nu | \Omega_\mu \rangle.
\end{equation}
The transition probability to obtain the output mode occupation $\vec{S}$ from the input mode occupation $\vec{R}$ is then obtained by \cite{Dittel2019}
\begin{equation}
    \label{eq:output_probability}
    p_{\vec{R}\rightarrow\vec{S}}=S\sum_{\mu,\nu\in\Sigma}[\rho_\mathrm{E}]_{\mu,\nu}\prod_{\alpha=1}^{N}
    \mathcal{U}_{F_\alpha,E_{\mu(\alpha)}} \mathcal{U}^*_{F_\alpha,E_{\nu(\alpha)}},
\end{equation}
with $S=N!/|S_{\vec{S}}|$, $|S_{\vec{S}}|=\prod_{j=1}^{n} S_j!$ and $\vec{F}$ the output mode assignment list (defined similar to the input mode assignment list $\vec{E}$).
With Eq.~\eqref{eq:output_probability}, we can calculate all relevant input-output probabilities for each input state listed in Table \ref{tab:gen_prob}. Note that 
an output fourfold coincidence can originate from any of the four-photon input states $\vec{R}_1$, $\vec{R}_2$, $\vec{R}_3$ with no photon being lost. However, fourfold coincidences can also arise from a six-photon input state with up to two photons ending up in ancillary modes and/or up to two photons lost after the unitary and/or up to three photons occupying the same output mode (collision event). All these cases are subsumed as six-photon background in Fig.~\ref{fig:results}.

\section{\label{sec:appendix:unitary_reconstruction} Waveguide fabrication and unitary characterization}

\begin{figure*}
\includegraphics[width=\linewidth]{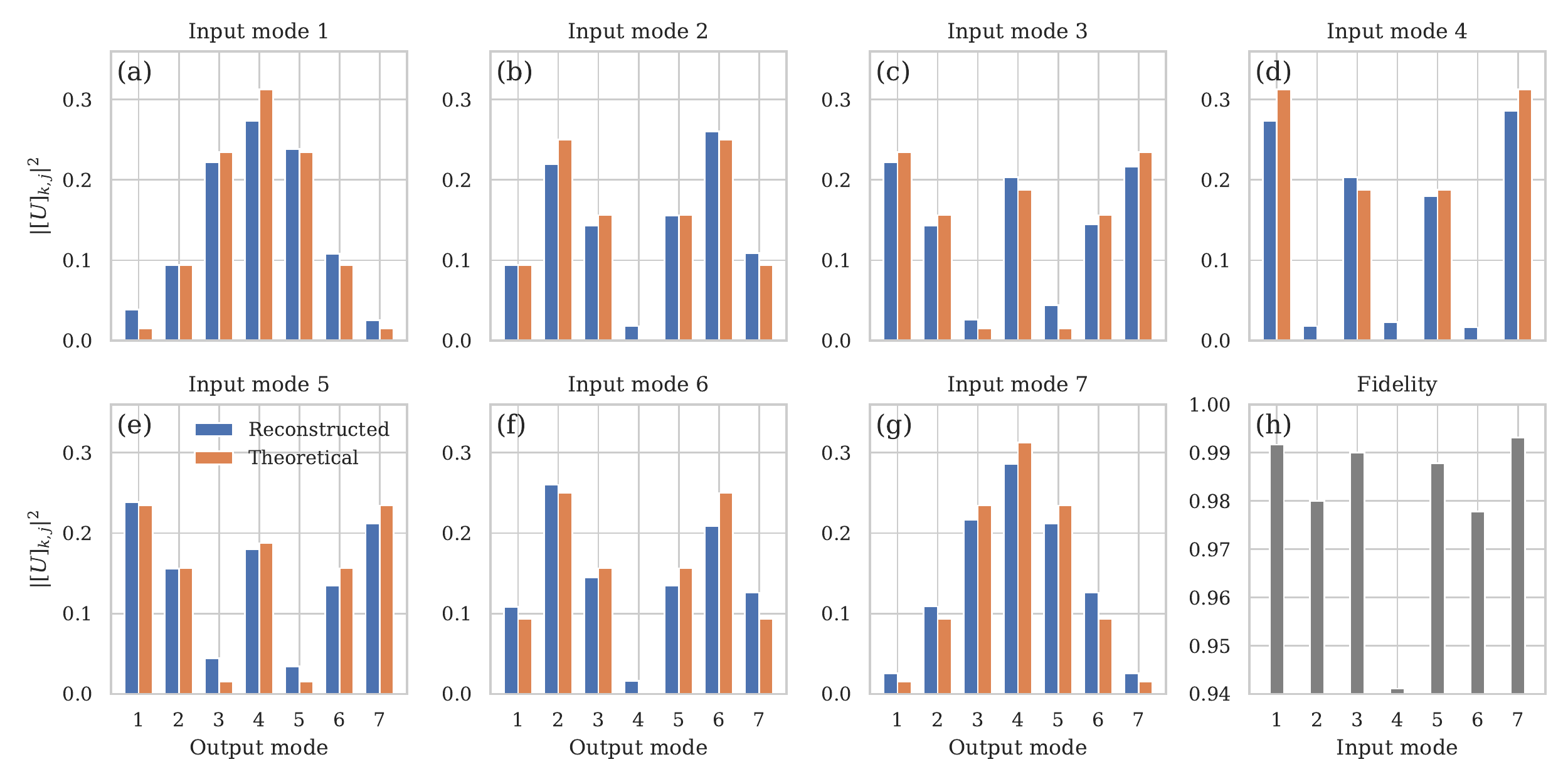}
\caption{\label{fig:jx-unitary} Reconstructed amplitudes of the $J_x$ unitary. We launch single photons into each input mode of the waveguide structure and measure the relative output intensity in each output mode. From this, we reconstruct the amplitudes of the $J_x$ unitary, $|[U_\mathrm{recon}]_{k,j}|^2$, following the algorithm from Ref.~\cite{Meany2012}. Panel (h) shows the fidelities $F_j=\left(\sum_{k=1}^{7}\sqrt{|[U_{\mathrm{recon}}]_{k,j}|^{2}\cdot |[U^\mathrm{J}]_{k,j}|^{2}}\right)^2$ of the reconstructed amplitudes, which benchmark how well the physical unitary agrees with the ideal $J_x$ unitary.
}
\end{figure*}

We fabricated the waveguide chip implementing the $J_x$ unitary in fused silica (Corning 7980, ArF grade) using the femtosecond laser direct-write approach \cite{Szameit2010}. Due to nonlinear absorption, the transparent material is modified within the focal region, producing a local increase of the refractive index. A Coherent RegA 9000 amplifier seeded by a Mira Ti:Sa femtosecond laser oscillator was used. After amplification, the pulses with \SI{150}{fs} pulse duration centred at \SI{800}{nm} had an energy of \SI{450}{nJ} at a repetition rate of \SI{100}{kHz}. Waveguides were permanently inscribed in the bulk while moving the sample at a constant speed of \SI{60}{mm\, min^{-1}} (high-precision positioning stages ALS 130, Aerotech Inc.~with a positioning error of $\pm\SI{0.1}{\micro m}$). The beam was focused with a $20\times$ objective, producing a mode field diameter of the guided mode of the order of \SI{18}{\micro m} x \SI{20}{\micro m} at \SI{815}{nm}. Fan-in and fan-out sections were arranged in a three-dimensional geometry 
to permit coupling to fiber arrays with standard spacing of \SI{127}{\micro m} while minimizing cross-talk.

Prior to the experiment, we characterized the unitary using single photons. The experimentally reconstructed amplitudes of the unitary matrix are compared to the ideal amplitudes of the $J_x$ unitary in Fig.~\ref{fig:jx-unitary}.
We reconstructed the unitary by sending single photons from the \ac{spdc} source into each input mode separately. For each input, we monitored the single-photon count rate in each of the seven output modes using \acp{apd}. By performing a least squares optimization, we reconstructed the amplitudes $|[U_{\mathrm{recon}}]_{k,j}|^2$ according to the procedure in \cite{Meany2012}. From this data we additionally retrieved the relative output transmissivities (including detector efficiencies), which we used in our simulation of the four-photon output probabilities (theory bars in Fig.~\ref{fig:results}). The phases of the matrix elements, $[\mathrm{Arg}(U_\mathrm{recon})]_{k,j}$,  were not reconstructed, however, from our experience with similar waveguide structures 
we expect that the fabricated structure closely matches the ideal phases of the $J_x$-lattice. According to experience, the propagation constant of the laser-written waveguides (which ultimately determines this phase) is much less sensitive to fabrication errors than the coupling rate between neighboring waveguides (which only influences the amplitudes of the unitary matrix). A two-photon interference experiment conducted on the sample supports our hypothesis, since its results agree well with the theoretical predictions, which are based on the ideal phases of the $J_x$ unitary.
Therefore, we model the matrix elements of the unitary probed in our experiment as $[U_{\mathrm{exp}}]_{k,j}=|[U_{\mathrm{recon}}]_{k,j}|\cdot e^{\mathrm{i}[\mathrm{Arg}(U^\mathrm{J})]_{k,j}}$.

\section{\label{sec:appendix:measurement_procedure} Procedure to temporally overlap the \texorpdfstring{\ac{spdc}}{SPDC} channels}

We adjusted the optical delays between the four inputs to the unitary by performing two-photon \ac{hom}-type measurements on the waveguide structure. First, we launched channels c and d into input modes 3 and 5, respectively. We monitored coincidences between output modes 1 and 4, while scanning the optical delay stage. The corresponding output state is almost perfectly suppressed for indistinguishable photons, while showing a reasonably high output probability for distinguishable photons. 
In the same fashion, we optimize source channel b relative to channel d, and thereafter channel a relative to channel b, while monitoring for each delay scanning multiple output combinations at the same time that exhibit good contrast in the detected coincidence rate between distinguishable and indistinguishable photons. 

\section{\label{sec:appendix:phase_fluctations} Phase fluctuation from twofold coincidence data}

\begin{figure}
\includegraphics[width=\linewidth]{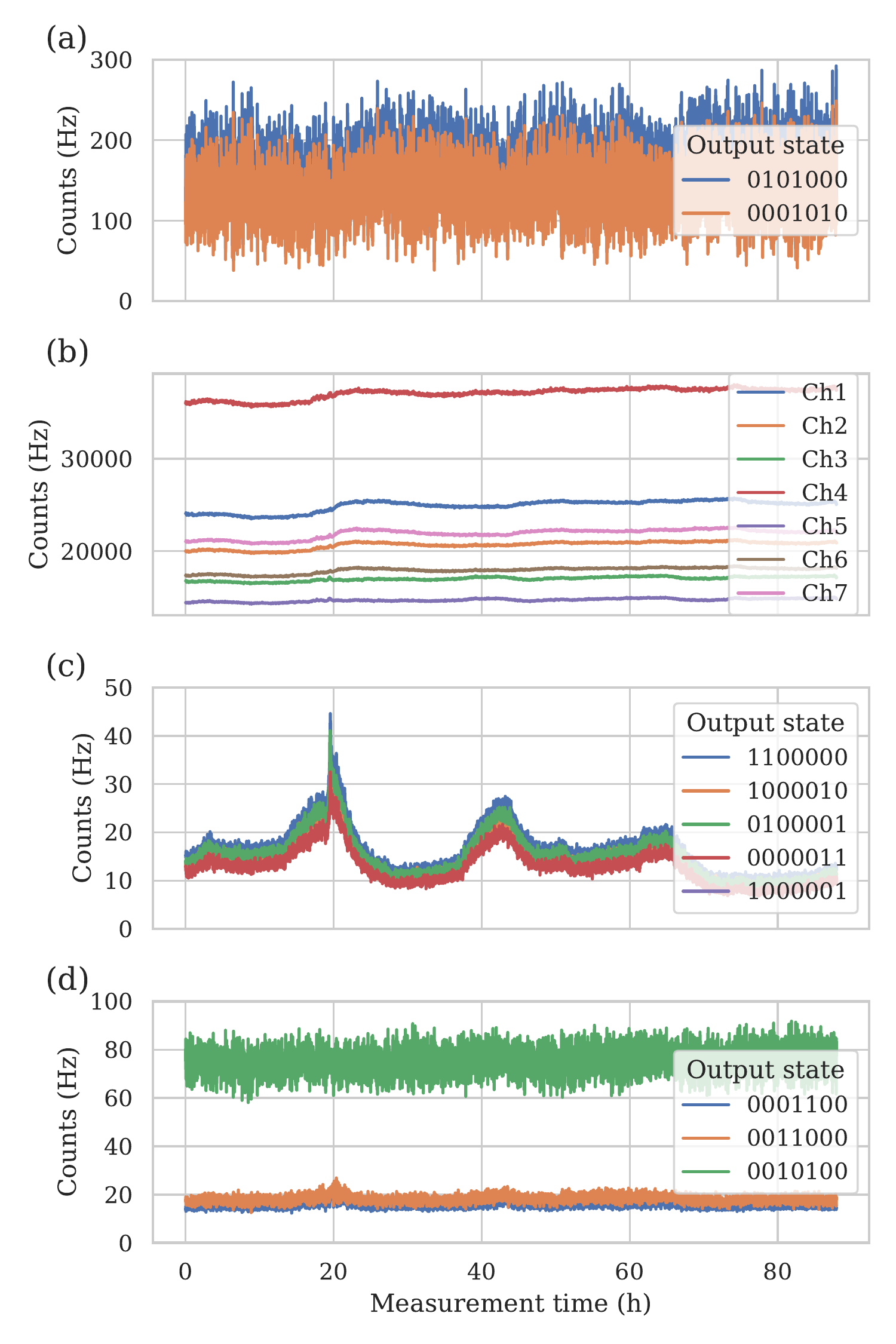}
\caption{\label{fig:time_traces} Selected two-fold coincidence counts and single-photon count rates during the measurement of the \emph{mutually indistinguishable} particle scenario. Counts are averaged over an interval of $\SI{60}{s}$ for a total measurement time of 88 hours. 
(a) Time traces of two output states which have approximately equal contribution from input state $\vec{R_4}=(0,0,1,0,1,0,0)$ and $\vec{R_5}=(1,0,0,0,0,0,1)$ according to the theoretical predictions.
The observed coincidence rate fluctuates heavily suggesting that the relative phase of these two contributions fluctuates.
(b) Time traces of single-photon count rates of all seven output channels.
(c) Selected time traces of output states which are dominated by contributions of input state $\vec{R_4}$. These output states are ideally suppressed according to the suppression law. We attribute the slow drift in the observed coincidence rates to a change in the indistinguishability of the photons caused by changes of the optical path length. Almost no fast fluctuations in the coincidence rate are observed since mostly one input state contributes. 
(d) Selected time traces of output states, which are dominated by contributions of input state $\vec{R_5}$. Neither strong fast fluctuations nor significant slow drifts of the coincidence rates are observed. 
}
\end{figure}

In general, one needs to take into account the relative phases of all contributing input states in the calculation of the final output probability distribution.
By applying and averaging over a series of phases in one of the input modes one can realize a mixed state, which allows for simpler addition of probabilities. This can be realized via a combination of quarter-, and half-wave plates oriented such that the input state is kept unchanged except for an additional phase factor \cite{Carolan2014, Jones2020}. 
In our experiment the mixed state is automatically obtained via time integration, as the relative phases fluctuate rapidly during the measurement.
This can be shown by looking at the time traces of twofold coincidence counts for the scenario of \emph{mutually indistinguishable} photons. These time traces were recorded parallel to the four-photon coincidences of the main experiment. Figure~\ref{fig:time_traces}(a) depicts the coincidence rate of two output states with approximately equal contribution from input states $\vec{R_4}=(0,0,1,0,1,0,0)$ and $\vec{R_5}=(1,0,0,0,0,0,1)$
[see the two rightmost states in Fig.~\ref{fig:2folds_indist_opt}]. The coincidence rates fluctuate strongly from 50 to \SI{250}{Hz}, while the single-photon count rates keep approximately constant [Fig.~\ref{fig:time_traces}(b)]. This suggests that the relative phase of the contributing inputs fluctuated, e.g.~through temperature drifts of the optical fibers from the source to the unitary.

\section{\label{sec:appendix:hom_vis} Estimation of average HOM visibility from twofold coincidences and JSA reconstruction}

\begin{figure}
\includegraphics[width=\linewidth]{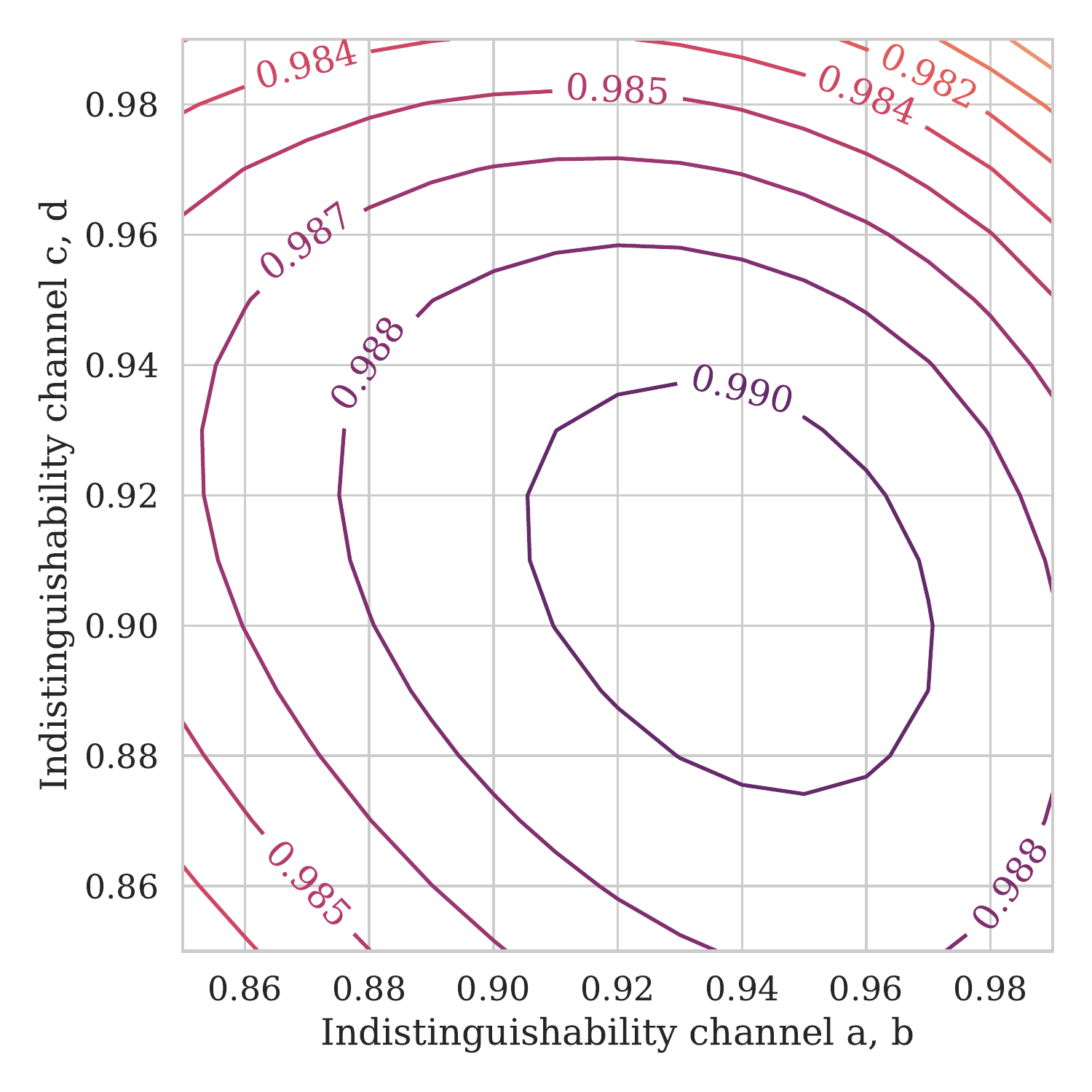}
\caption{\label{fig:fidelity_2fold} Calculated fidelity for an output probability distribution of all twofold, non-bunching coincidence events as a function of the indistinguishabilities $V_{\mathrm{ab}}$ and $V_{\mathrm{cd}}$.}
\end{figure}

\begin{figure}
\includegraphics[width=\linewidth]{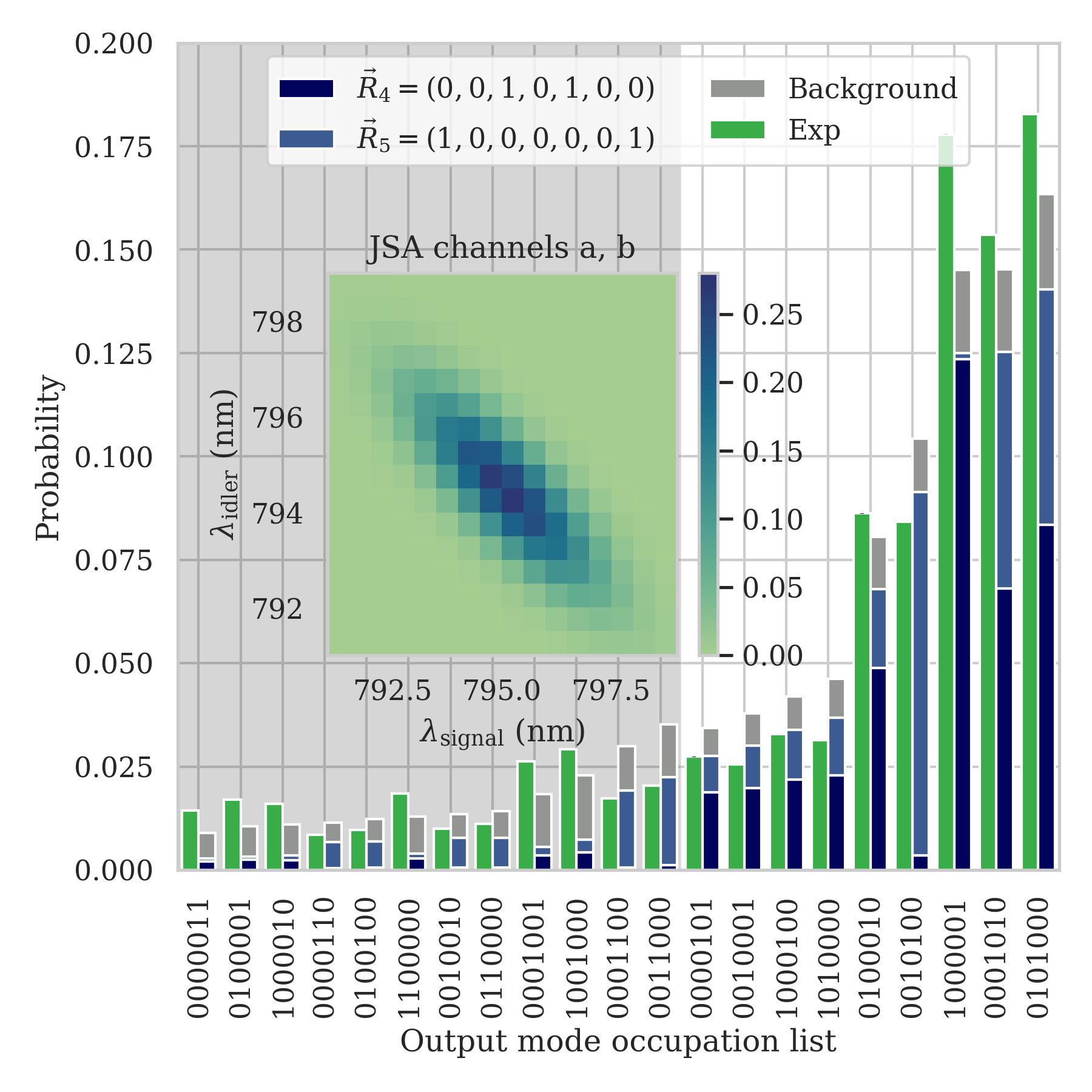}
\caption{\label{fig:2folds_indist_opt} Output probability distribution of all twofold, non-bunching coincidence events. For the simulated result, we used optimized time-averaged indistinguishabilities of $V_{\mathrm{ab}}=0.94$ and $V_{\mathrm{cd}}=0.90$. The fidelity is $F=0.991$. The experimental and simulated degree of suppression violation are $D_\mathrm{exp}=0.198$ and $D_\mathrm{sim}=0.202$, respectively. Experimental error bars from the counting statistics are too small to be visible. The inset shows the discretized \ac{jsa} of channels a and b (the \ac{jsa} of channels c and d is similar).}
\end{figure}

For the investigated scenario of \emph{mutually indistinguishable} particles we observed a slow change in the recorded twofold coincidence counts over time [compare Fig.~\ref{fig:time_traces}(c) and (d)]. 
This effect is most pronounced for output states, which receive contributions mostly from the input state $\vec{R_4}$.
We attribute this drift to relative optical path length changes in the pertinent channels (here c and d) of the setup from the \ac{spdc} source to the $J_x$ unitary, which leads to a degradation of the indistinguishability from the initially optimized situation. This may arise from the asymmetric response to temperature changes of the freespace delay in channel c and the fiber in channel d, and changes the output probability distribution (counting statistics), e.g.~it may increase the probability to observe an ideally suppressed output event due to a degrading of the particles' indistinguishability.

We estimate the time-averaged degraded indistinguishability of photons from the same pair from the total experimental twofold coincidence counts, which we collect in parallel to the fourfold coincidences of the main experiment. We compare the experimentally obtained output distribution of all non-bunching two-photon events with the simulated output distribution for varying indistinguishabilities $V_{\mathrm{ab}}$ and $V_{\mathrm{cd}}$, i.e., indistinguishability between photons in channels a and b, as well as channels c and d. Additionally, we keep the indistinguishability between photons originating from different pairs at a fixed value of 59.1\% (retrieved from a separate heralded \ac{hom} measurement, see Appendix \ref{sec:appendix:source_setup_parameters}). 

For each simulated output distribution, we calculate the fidelity according to 
\begin{equation}
    F=\left(\sum_i\sqrt{p_{\mathrm{exp},i}\,p_{\mathrm{sim},i}}\right)^2,
\end{equation}
with the sum running over all non-bunching two-photon outputs.
We consider input states $\vec{R_4}$ and $\vec{R_5}$, as well as a background of four- and six-photon states, where two and four photons are lost before the unitary, respectively (these are the dominant contributions under the considerations of losses). 
The background resulting from initial states of four- and six-photons depends on the indistinguishability between photons from different pairs and contributes around 22.5\% to the measured final output events. 
As shown in Fig.~\ref{fig:fidelity_2fold}, the fidelity reaches its maximum value for indistinguishabilties of $V_{\mathrm{ab}}=0.94$ and $V_{\mathrm{cd}}=0.90$. These values should therefore best represent the time-averaged indistinguishabilties in the experiment.
For the above estimated time-averaged indistinguishabilities, the normalized output distribution of all non-bunching two-fold coincidence events is plotted in Fig.~\ref{fig:2folds_indist_opt}. It reaches a fidelity of $F=0.991$.

We finally calculate a \ac{jsa} that reproduces the measured and estimated indistinguishabilities by following the procedure in \cite{Kolenderski2009}. First a phase matching function is calculated in the paraxial approximation of the involved spatial modes of the pump laser and the photon-collecting fibers. We apply a Gaussian pump spectral envelope and spectral filter functions (\SI{3}{nm} FWHM bandwidth, as in the setup) to obtain a discretized \ac{jsa} on a frequency grid of size $17\times17$. The \ac{jsa} of \ac{spdc} channel a and b is plotted in the inset of Fig.~\ref{fig:2folds_indist_opt}. 

Instead of modelling the reduction of indistinguishability by a path-length drift directly in the time-domain, we mimick this effect by mutually shifting the central transmission frequencies of the filters 
(thus reducing spectral instead of temporal overlaps).
The estimated time-averaged indistinguishabilities are reproduced with an up-converted pump spectral width of \SI{0.4}{nm} FWHM, as well as pairwise filter offsets of \SI{0.625}{nm} between channels a and b and \SI{0.8}{nm} between channels c and d. 
The internal state of multiple pairs is then calculated as a tensor product of the two-photon \acp{jsa} [cf. Eq.~\eqref{eq:internalstate_multipairs}].

\section{\label{sec:appendix:source_setup_parameters} Source parameters and photon loss}

To properly simulate the source  and subsequent channel losses in the experiment, we needed to estimate the pair generation probability, the collection efficiencies of the source, as well as the indistinguishabilities of photons from different channels.

A preliminary characterisation of the source lets us obtain the pair generation probability and the collection efficiencies of the four source channels. We attached each source channel directly to an \ac{apd}, which resulted in typical single-photon count rates of $160-\SI{370}{kHz}$ at a pump power of $\approx\SI{150}{mW}$. Additionally, we measured coincidence rates of $\SI{31}{kHz}$ between channels a and b, as well as $\SI{25}{kHz}$ between channels c and d. From these rates, we reconstructed pair generation probabilities per pump pulse of $p^{(\mathrm{ab})}=2.6\%$ and $p^{(\mathrm{cd})}=3.3\%$ for channels a and b, and channels c and d, respectively. Additionally, we obtained collection efficiencies of 19\%, 18\%, 10\%, and 22\% for the four channels a-d. In the source characterization, we used detectors with a detection efficiency of 68\%. The collection efficiencies were corrected for this detection efficiency. 

These channel collection efficiencies as well as the incoupling efficiencies to the seven input modes of the unitary, which includes fiber-to-chip coupling and propagation loss on the chip, are combined to input mode transmissivity factors $\eta_j$ for $j=1,\ldots,7$. We extracted the incoupling efficiencies from the unitary-reconstruction data [cf. Appendix \ref{sec:appendix:unitary_reconstruction}], comparing the total transmissivity across all output modes for the relevant input modes and assuming an on-chip propagation loss of $\SI{0.5}{dB/cm}$ in all straight parts of the waveguide structure. 
For the four occupied input modes in the experiment, we calculate input mode transmissivity factors $\eta_1=0.055$, $\eta_3=0.034$, $\eta_5=0.107$, and $\eta_7=0.065$. 

We performed a heralded \ac{hom}-interference experiment on a fiber beamsplitter, where we interfered photons from channels b and d, using a and c as additional heralding channels \cite{Ou1999a}.
From the experiment, we attained an indistinguishability of 59.1\% for photons originating from different pairs.
In a standard \ac{hom} experiment with photons from the same pair, we retrieved maximal visibilities of 96.1\% for source channels a and b, and 98.2\% for source channels c and d. In the actual experiment, these indistinguishabilties vary over time, probably due to optical path length drifts, which reduces the temporal overlap of photons from different channels [see Appendix \ref{sec:appendix:hom_vis}]. 


\FloatBarrier 

\bibliography{ms} 

\end{document}